\documentclass[%
reprint,
superscriptaddress,
 amsmath,amssymb,
 aps,
prc,
]{revtex4-1}

\usepackage{graphicx}
\usepackage{dcolumn}
\usepackage{bm}
\usepackage{color}

\def\be{\begin{eqnarray}}   
\def\ee{\end{eqnarray}}

\newcommand{\affA}{
Max Planck Institute for the Structure and Dynamics of Matter and Center for Free-Electron Laser  Science, Luruper Chaussee 149, 22761 Hamburg, Germany
}
\newcommand{\affB}{
Department of Physics, Politecnico di Milano, Piazza Leonardo da Vinci 32, 20133 Milano, Italy
}
\newcommand{\affC}{
Department of Physics, ETH Z\"urich, 8093 Z\"urich, Switzerland
}
\newcommand{\affD}{
Center for Computational Quantum Physics (CCQ), The Flatiron Institute, 162 Fifth avenue, New York NY 10010
}

\begin{document}

\preprint{APS/123-QED}

\title{Role of intra-band transitions in photo-carrier generation}

\author{Shunsuke~A.~Sato}\affiliation{\affA}
\author{Matteo~Lucchini}\affiliation{\affB}
\author{Mikhail~Volkov}\affiliation{\affC}
\author{Fabian~Schlaepfer}\affiliation{\affC}
\author{Lukas~Gallmann}\affiliation{\affC}
\author{Ursula~Keller}\affiliation{\affC}
\author{Angel~Rubio}\affiliation{\affA}\affiliation{\affD}

\date{\today}

\begin{abstract}
We theoretically investigate the role of intra-band transitions in laser-induced
carrier-generation for different photon energy regimes:
(i) strongly off-resonant, (ii) multi-photon resonant, and (iii) resonant conditions.
Based on the analysis for the strongly off-resonant
and multi-photon resonant cases,
we find that intra-band transitions strongly enhance photo-carrier generation
in both multi-photon absorption and tunneling excitation regimes,
and thus, they are indispensable for describing the nonlinear photo-carrier generation processes.
Furthermore, we find that intra-band transitions
enhance photo-carrier generation even in the resonant condition,
opening additional multi-photon excitation channels once the laser irradiation
becomes sufficiently strong.
The above findings suggest a potential for efficient control 
of photo-carrier generation via multi-color laser pulses through optimization of the  
contributions from intra-band transitions.
\end{abstract}

\maketitle


\section{Introduction \label{sec:intro}}

Thanks to the technological advances in the generation of ultrashort 
laser pulses over the last decades, time-domain electron dynamics in atoms and molecules became accessible
\cite{RevModPhys.81.163,goulielmakis2010real,Schultze1658,warrick2016probing,Isinger893}. 
Recently, the time-domain observation technique
has been further applied to solid-state materials, and laser-induced 
ultrafast electron dynamics in solids has been intensively investigated
\cite{Science.346.1348,Science.353.6302,zurch2017direct}.
One of the most characteristic features of solid-state systems, compared with atoms
and molecules, is the formation of continuum energy bands due to the large number of 
electrons in the systems. The continuum energy bands enable transitions within
each band in addition to transitions among different bands.
The first one is called \textit{intra-band} transition, 
while the latter one is called \textit{inter-band} transition.
The importance of intra-band transitions under intense laser fields has been extensively
discussed in the context of high-order harmonic generation from solids 
\cite{ghimire2011observation,vampa2015linking,luu2015extreme,PhysRevLett.113.073901,PhysRevX.7.021017}.
However, their role is still unclear and under debate.
Furthermore, it has been demonstrated that intra-band processes play a significant role
in the ultrafast modification of optical properties of dielectrics under intense 
laser fields in various conditions \cite{Science.353.6302,PhysRevLett.117.277402}.
Recently, the pivotal role of intra-band transitions in photo-carrier generation
of GaAs was pointed out by Schlaepfer, \textit{et al} \cite{Schlaepfer2017}.

Photo-carrier generation is one of the most fundamental processes in laser-solid
interaction as it triggers various phenomena such as band-gap renormalization
\cite{PhysRevLett.58.419,PhysRevB.41.8288},
formation of electron-hole plasmas \cite{PhysRevB.61.2643}, and laser ablation
\cite{gattass2008femtosecond,bonse2017laser}.
The theoretical description of photo-carrier generation under strong fields
has been studied for a long time 
\cite{keldysh1965,PhysRevB.75.205106,PhysRevA.96.063410,McDonald2017,PhysRevLett.118.173601}.
One of the most successful approaches
is the Keldysh theory \cite{keldysh1965}, which finds application in many fields
\cite{balling2013femtosecond,PhysRevLett.82.3883,doi:10.1117/12.2244833}. 
Although the Keldysh theory and its modifications
succeed to describe the photo-carrier injection rate fairly well, 
the role of intra-band transitions in the photo-carrier generation process remained unclear.
Therefore, further understanding of intra-band transitions 
is important to find a way to enhance or suppress photo-carrier generation
under rather complex conditions such as complicated materials as well as
multi-color laser fields, which are out of the scope of the Keldysh theory.

In this work, we theoretically investigate the role of intra-band transitions 
in photo-carrier generation with a parabolic two-band model, which
is the simplest model to describe semiconductors and insulators.
The model has been successfully applied to investigate static as well as dynamical optical
properties \cite{yu.cardona2010,PhysRevLett.76.4576,PhysRevB.93.045124}.
To explore the carrier-generation processes, 
we calculate the number of laser-induced carriers
with the two-band model in three photon-energy regimes: 
(i) strongly off-resonant, (ii) multi-photon resonant, and (iii) resonant excitation.
In each regime, we analyze the effects of intra-band transitions
on photo-carrier generation by artificially suppressing
intra-band transitions in the two-band model.
As a result of the analysis, the impact of intra-band transitions
will be isolated for each regime.

The paper is organized as follows: In Sec. \ref{sec:model}
we first describe the parabolic two-band model, which will be used in our analysis.
Then, we demonstrate the accuracy of the model by comparing it with Kane's band model 
\cite{j.phys.chem.sol.1.249} 
as well as \textit{ab-initio} simulations.
In Sec. \ref{sec:intra} we investigate the role of intra-band transitions
in photo-carrier generation, computing the electron dynamics under laser fields
with the two-band model.
Finally, our findings are summarized in Sec. \ref{sec:summary}.

\section{Theoretical model \label{sec:model}}

In this section, we briefly introduce the parabolic two-band model that will be used 
in this work. Then, we demonstrate the accuracy of the model for the nonlinear electron
dynamics under strong fields by comparing it with
the non-parabolic Kane's band model \cite{j.phys.chem.sol.1.249} 
and \textit{ab-initio} calculations 
based on the time-dependent density functional theory (TDDFT) \cite{PhysRevLett.52.997}.

\subsection{Parabolic two-band model \label{subsec:2-band-model}}

To construct the parabolic two-band model, we start from the following one-body 
non-relativistic Schr\"odinger equation in the dipole approximation
\be
i\hbar \frac{\partial}{\partial t}u_{b\vec k}(\vec r,t)
&=&\left [ \frac{1}{2m_e}
\left \{ \vec p + \hbar \vec k + \frac{e}{c}\vec A(t) \right \}^2
 + v(\vec r)\right ]  u_{b\vec k}(\vec r,t) \nonumber \\
&=&\hat h_{\vec K(t)} u_{b\vec k}(\vec r,t),
\label{eq:schrodinger-full}
\ee
where $m_e$ is the electron mass, $u_{b\vec k}(\vec r,t)$ is a time-dependent Bloch state, 
and $v(\vec r)$ is a one-body potential that has the same periodicity as the crystal.
Here $b$ denotes a band index, while $\vec k$ denotes the Bloch wave number.
We note that the crystal momentum is shifted by the vector potential as
$\vec K(t) = \vec k + e\vec A(t)/\hbar c$ based on the acceleration theorem.

Then, we introduce the Houston state \cite{PhysRev.57.184,PhysRevB.33.5494} as
\be
u^H_{b\vec k}(\vec r,t) = \exp \left [
-\frac{i}{\hbar}\int^t dt' \epsilon_{b\vec K(t')} 
\right ] 
u^{S}_{b\vec K(t)}(\vec r),
\label{eq:houston-state}
\ee
where $\epsilon_{b\vec K(t)}$ and $u^{S}_{b\vec K(t)}(\vec r)$ are 
an eigenvalue and the eigenstate of the instantaneous Hamiltonian, $\hat h_{\vec K(t)}$,
respectively;
\be
\hat h_{\vec K(t)} u^{S}_{b\vec K(t)}(\vec r) 
= \epsilon_{b\vec K(t)} u^{S}_{b\vec K(t)}(\vec r).
\label{eq:inst-eigenstate}
\ee

To construct a two-band model, we assume that the wave-function at each $k$-point 
can be expanded by only two Houston states; one representing a valence, and the other a conduction
state;
\be
u_{\vec k}(\vec r,t) = c_{v\vec k}(t) u^H_{v\vec k}(\vec r,t)
+c_{c\vec k}(t) u^H_{c\vec k}(\vec r,t).
\label{eq:wf-expand-houston}
\ee

Inserting Eq. (\ref{eq:wf-expand-houston}) into Eq. (\ref{eq:schrodinger-full}),
one can derive an equation of motion for the coefficients $c_{v\vec k}(t)$ 
and $c_{c\vec k}(t)$,

\be
i\hbar
\frac{d}{dt}
\left(
    \begin{array}{c}
      c_{v\vec k}(t)   \\
      c_{c\vec k}(t)
    \end{array}
  \right)
=
\left(
    \begin{array}{cc}
      0 & h_{vc,\vec k}(t)  \\
      h^*_{vc, \vec k}(t) & 0
    \end{array}
  \right)
\left(
    \begin{array}{c}
      c_{v\vec k}(t)   \\
      c_{c\vec k}(t)
    \end{array}
  \right),
\nonumber \\
\label{eq:schrodinger-2band}
\ee
where the off-diagonal matrix element is given by
\be
h_{vc,\vec k}(t) = -\frac{i \vec p_{vc,\vec K(t)} \cdot \vec E(t) }
{\epsilon_{v, \vec K(t)} -\epsilon_{c, \vec K(t)}}\frac{e \hbar}{m}
e^{\frac{1}{i\hbar}\int^t dt' \left \{
\epsilon_{c, \vec K(t')} - \epsilon_{v, \vec K(t')}
\right \}
},
\nonumber \\
\ee
and
\be
\vec p_{vc,\vec K(t)} = \int_{\Omega} d\vec r
u^{S,*}_{v\vec K(t)}(\vec r) \vec p u^{S}_{c\vec K(t)}(\vec r),
\ee
where $\Omega$ is the volume of the unit-cell.
Note that Eq.~(\ref{eq:schrodinger-2band}) is nothing but 
the Houston state expansion of the Schr\"odinger equation
\cite{PhysRev.57.184,PhysRevB.33.5494} with only two Houston states.

To further simplify the model, we apply two approximations:
(i) parabolic band approximation, and (ii) uniform matrix-element approximation.
The parabolic band approximation allows us to describe the electronic structure 
by the following quadratic forms:
\be
\epsilon_{v,\vec k} &=& - \frac{\hbar^2 \vec k^2}{2m_{v}}, \\
\epsilon_{c,\vec k} &=& \epsilon_g + \frac{\hbar^2 \vec k^2}{2m_{c}},
\ee
where $\epsilon_g$ is the band gap, and $m_v$ and $m_c$ are the effective masses for
valence and conduction bands, respectively. The uniform matrix-element approximation allows us to ignore
the $k$-dependence of the matrix-element,
\be
\vec p_{vc,\vec K(t)} = \vec p_{vc}.
\ee
Note that the uniformity of the matrix elements has been confirmed
by \textit{ab-initio} simulations for several semi-conductors \cite{PhysRev.142.530}.

The derived parabolic two-band model is the simplest model
for semiconductors and insulators.
The accuracy of this model in the context of carrier generation
under strong fields will be evaluated in the following subsection.

\subsection{Comparison with other models \label{subsec:comp-tddft}}

To demonstrate the accuracy of the parabolic two-band model, 
we compare it with the non-parabolic Kane's band model
\cite{j.phys.chem.sol.1.249}
and first-principles calculations based on the TDDFT.

In this work, we consider the laser-induced electronic excitation in $\alpha$-quartz.
To describe $\alpha$-quartz, we set the effective mass 
$m_r=1/(m^{-1}_v+m^{-1}_c)$ to $0.4m_e$ 
and the band gap $\epsilon_g$ to $9.0$ eV
according to Ref. \cite{PhysRevB.87.115201}. 
The transition momentum matrix $p_{vc}$
is evaluated by Kane's formula \cite{j.phys.chem.sol.2.181},
\be
p_{vc} = \frac{1}{2}\sqrt{\frac{\epsilon_g}{m_r}}.
\ee

\begin{figure}[htbp]
\centering
\includegraphics[width=0.9\columnwidth]{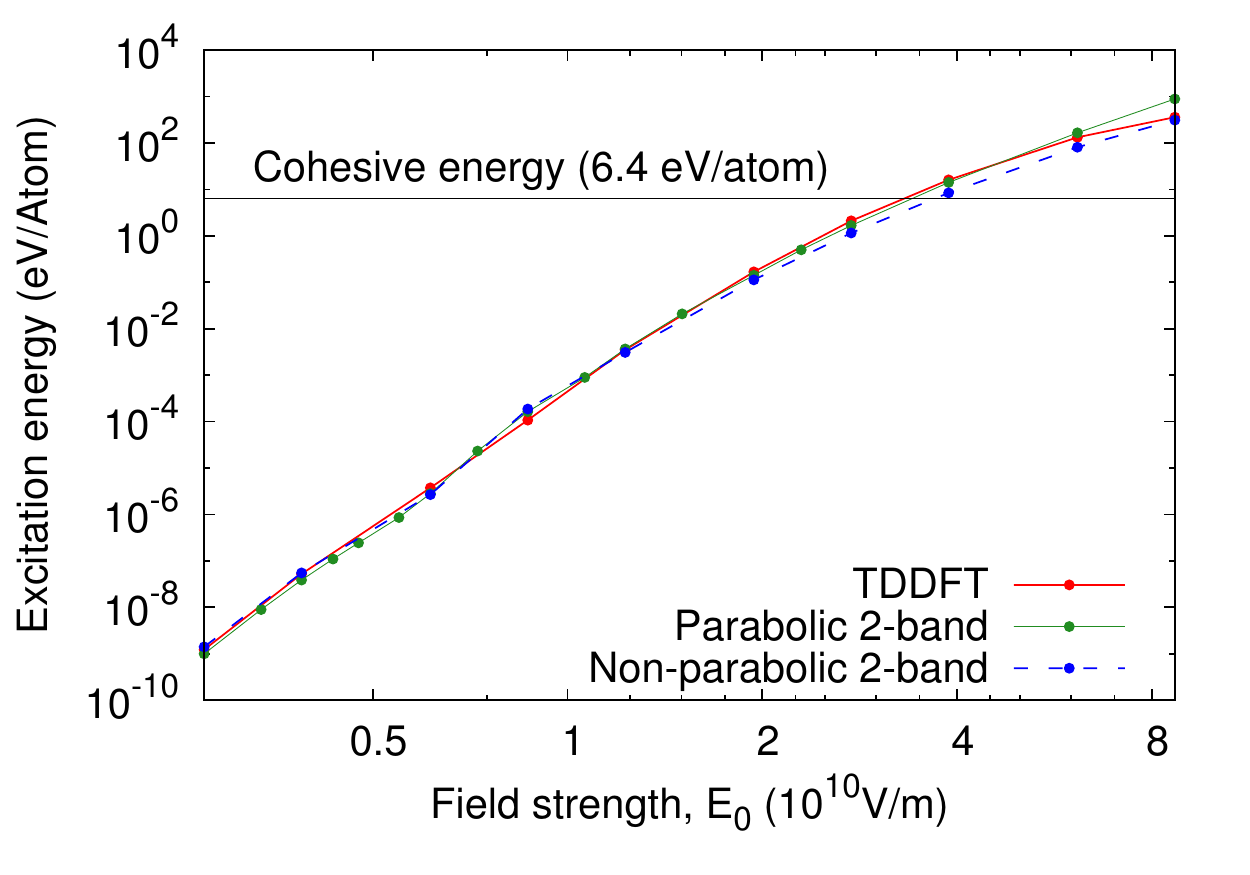}
\caption{\label{fig:SiO2_Eex_jule}Laser-induced electronic excitation energy
in $\alpha$-quartz after the laser irradiation.
The dots represent the results of the TDDFT calculation (red solid), 
the parabolic two-band model (green solid) and the non-parabolic two-band 
model (blue dashed).}
\end{figure}

The laser-induced electronic excitation energy $E_{ex}$ for the two-band model
is given by
\be
E_{ex} = \frac{2n_c}{(2\pi)^3}\int d\vec k \left ( 
\epsilon_{c\vec k} - \epsilon_{v\vec k}
\right ) |c_{c\vec k}(t_f)|^2,
\ee
where $t_f$ is the time at which the laser irradiation ends. Here, we introduced
a dimensionless factor $n_c$ so that 
the two-band model reproduces the first-principles calculation 
(see Fig.\ref{fig:SiO2_Eex_jule}). 
Since the electronic structure of the material is approximated by only the two bands,
the effective electron density that contributes to the response may not be 
well described. Thus, the factor $n_c$ can be understood as a correction
for the effective electron density.
In this work, we set $n_c$ to 20.

For the non-parabolic band model, we employ the following Kane's band 
\cite{j.phys.chem.sol.1.249}
instead of the parabolic band,
\be
\epsilon_{c \vec k} - \epsilon_{v \vec k}
= \epsilon_g \left (
1 + \frac{\hbar^2 k^2}{m_r\epsilon_g}
\right )^{1/2}.
\ee
We note that the Keldysh formula is derived based on this Kane's band model
\cite{keldysh1965}.

For the \textit{ab-initio} modeling, we employ real-time real-space TDDFT 
calculations with norm-conserving pseudopotential \cite{PhysRevB.43.1993}. 
For the exchange-correlation potentials,
the Becke-Johnson exchange \cite{JChemPhys.124.221101}
and Perdew-Wang correlation \cite{PhysRevB.45.13244} are employed.
For practical simulations, we employed a TDDFT code;
the \textit{ab-initio} real-time electron dynamics simulator (ARTED)
\cite{SA.Sato2014,Y.Hirokawa2016}.
Technical details of the real-time real-space TDDFT calculations are explained
elsewhere \cite{PhysRevB.62.7998,doi:10.1063/1.4937379}. Here, we only describe the numerical parameters for the TDDFT
calculation in this work: we employ a rectangular unit cell which contains six silicon atoms
and twelve oxygen atoms. The unit cell is discretized into a Cartesian $20\times36\times50$ grid.
For the Brillouin zone sampling, we employ $4^3$ $k$-points.

In this work, the applied electric field is described by the following vector potential:
\be
\vec A(t) = - \vec e_c \frac{E_0}{\omega} \sin^4\left( \pi \frac{t-\frac{T}{2}}{T}\right )
\sin \left(\omega \left (t-\frac{T}{2} \right) \right),
\nonumber \\
\ee
in the domain $0<t<T$ and zero outside.
Here, $\vec e_c$ is the unit vector for the $c$-axis of $\alpha$-quartz, 
$E_0$ is the maximum electric field strength, $T$ is the full pulse duration, 
and $\omega$ is the mean frequency. 

Here, we set the mean frequency $\omega$ to $1.55$ eV$/\hbar$, and 
the full duration $T$ to $30$ fs. We note that the corresponding full-width at half-maximum duration
of the laser intensity profile is about $7.8$ fs.
By changing the maximum field strength $E_0$, we compute 
the laser-induced electronic excitation energy.

Figure \ref{fig:SiO2_Eex_jule} shows the laser-induced electronic excitation energy calculated with
TDDFT (red), the parabolic two-band model (green), 
and the non-parabolic two-band model (blue).
The black horizontal line indicates the cohesive energy of $\alpha$-quartz, 
$6.4$~eV/atom \cite{weast1986handbook},
which is regarded as a reference of the laser-ablation threshold \cite{PhysRevB.92.205413}.
One can see that the parabolic two-band model provides almost the same result
as the non-parabolic Kane's band model. 
This fact indicates that the anharmonicity of the band structure does not have
significant effects on photo-carrier generation in the present conditions.

Furthermore, we emphasize that both the parabolic and 
the non-parabolic two-band models
show reasonable agreement with the first-principles results for the full intensity range studied here.
Therefore, the accuracy of this simplified parabolic
two-band model for the analysis of the laser-induced electronic excitation is demonstrated.

We note that, while the parabolic two-band model shows a nice agreement with 
the TDDFT calculation in the strongly off-resonant condition,
it may show deviations in other conditions due to a multi-band effect.
Although the multi-band effect is important to describe realistic systems,
we completely omit it in the present work in order to clearly understand 
the role of intra-band transitions.
The multi-band effect and its coupling with intra-band transitions will be investigated
in future work based on this work.

\section{Effects of intra-band transitions \label{sec:intra}}

\begin{figure}[htbp]
\centering
\includegraphics[width=0.9\columnwidth]{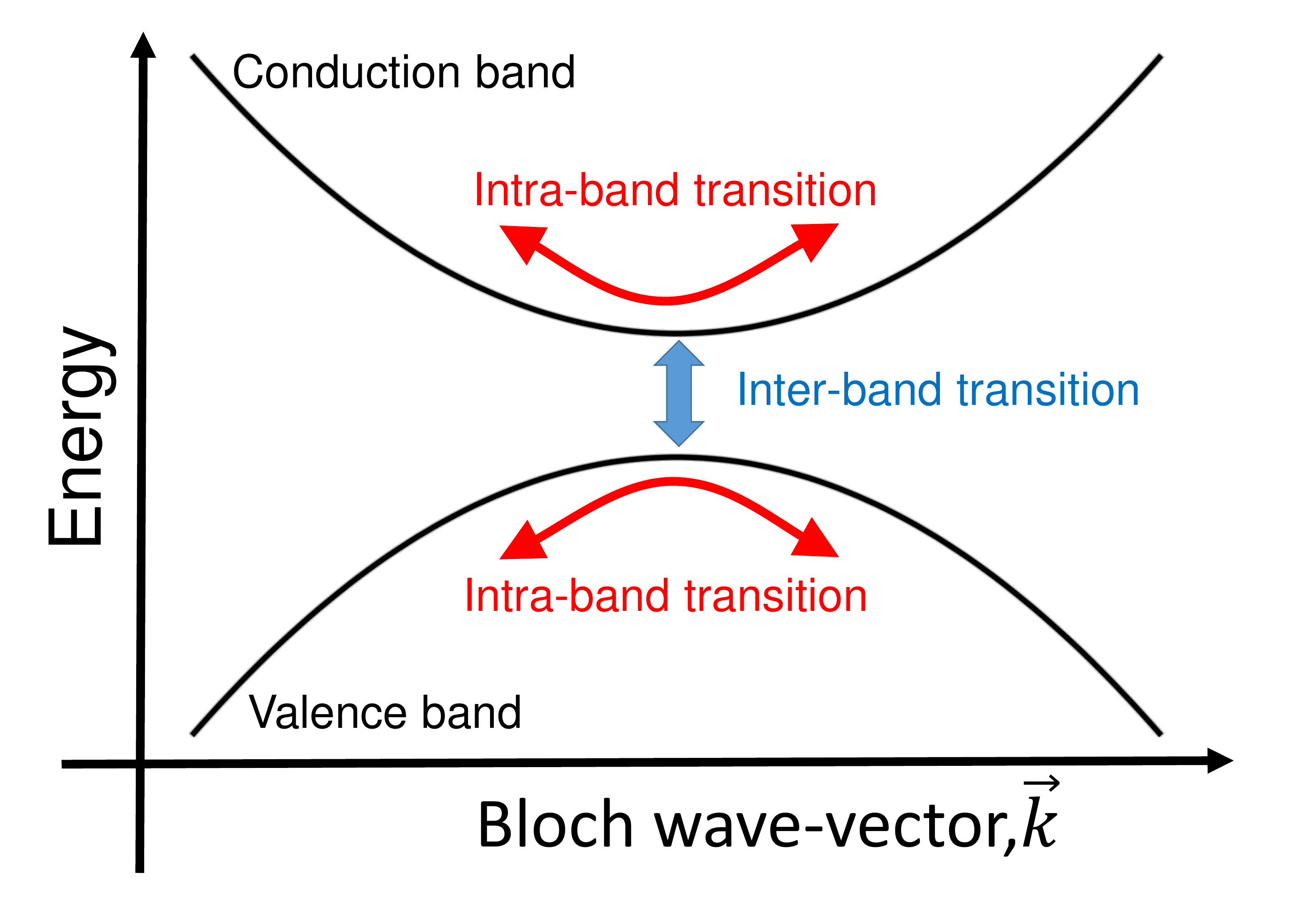}
\caption{\label{fig:two_band_fig} 
Schematic picture of intra-band transitions (red) and inter-band transitions (blue)
in the parabolic two-band model.
}
\end{figure}

In this section, we investigate the impact of intra-band transitions
on photo-carrier generation, employing the parabolic two-band model
described in the previous section.
Figure~\ref{fig:two_band_fig} shows a schematic picture of intra-band
and inter-band transitions; intra-band transitions correspond
to the horizontal motion in the $k$-space, while inter-band transitions
correspond to the vertical transition.

In the two-band model, intra-band transitions
are described by the shift of the crystal momentum $\vec K(t)$ 
in the instantaneous eigenstates in Eq. (\ref{eq:wf-expand-houston}), while 
inter-band transitions are described by transitions
between the valence and the conduction Houston states via the off-diagonal
elements of the Hamiltonian matrix.
Therefore, one can omit intra-band transitions from the model 
by neglecting the crystal momentum shift due to the vector potential:
$\vec K(t) = \vec k + e \vec A(t)/\hbar c \rightarrow \vec k$.
In order to investigate the impact of intra-band transitions, we compare 
the parabolic two-band model with (a) both intra and inter-band transitions enabled,
and (b) only inter-band transitions.
We shall call the first one \textit{full} model, and the latter one \textit{pure inter-band}
model.
A similar analysis has been done by Golde, \textit{et al.} for high-order harmonic
generation in a semiconductor using the semiconductor Bloch equations
\cite{PhysRevB.77.075330}.

\subsection{Strongly off-resonant condition; $\hbar \omega /\epsilon_g \sim 0.17$ 
\label{subsec:strong-off-resonant}}

First, we investigate the effect of intra-band transitions in the strongly off-resonant
regime, where the mean photon energy, $\hbar \omega = 1.55$~eV, is much smaller than 
the band gap, $\epsilon_g = 9$~eV. The full pulse duration is set to $T=30$~fs.
The corresponding bandwidth of the laser pulse is about $0.5$ eV.
It is well known that photo-carrier generation in such a strongly off-resonant regime
is well described by Keldysh's formula \cite{keldysh1965} with 
the so-called Keldysh parameter,
\be
\gamma = \frac{\omega \sqrt{m_r \epsilon_g}}{eE_0} = \frac{1}{2} 
\sqrt{\frac{\epsilon_g}{U_p}},
\ee
where $U_p$ is the ponderomotive energy; $U_p=e^2 E^2_0/4m_r\omega^2$.
The Keldysh parameter is commonly used to determine the excitation regime \cite{J.Phys.B.25.4005}:
a large value ($\gamma \gg 1$) indicates that multi-photon absorption dominates
the excitation mechanism, while a small value ($\gamma < 0.5$)
indicates that the tunnel mechanism dominates.
One may expect that intra-band transitions play 
a significant role in photo-carrier generation in the strongly off-resonant regime
since the ponderomotive energy $U_p$, which originates from the quiver motion
of the crystal momentum $\vec K(t)$ due to intra-band transitions, 
is linked to the excitation rate through $\gamma$ in the Keldysh theory \cite{keldysh1965}.

\begin{figure}[htbp]
\centering
\includegraphics[width=0.9\columnwidth]{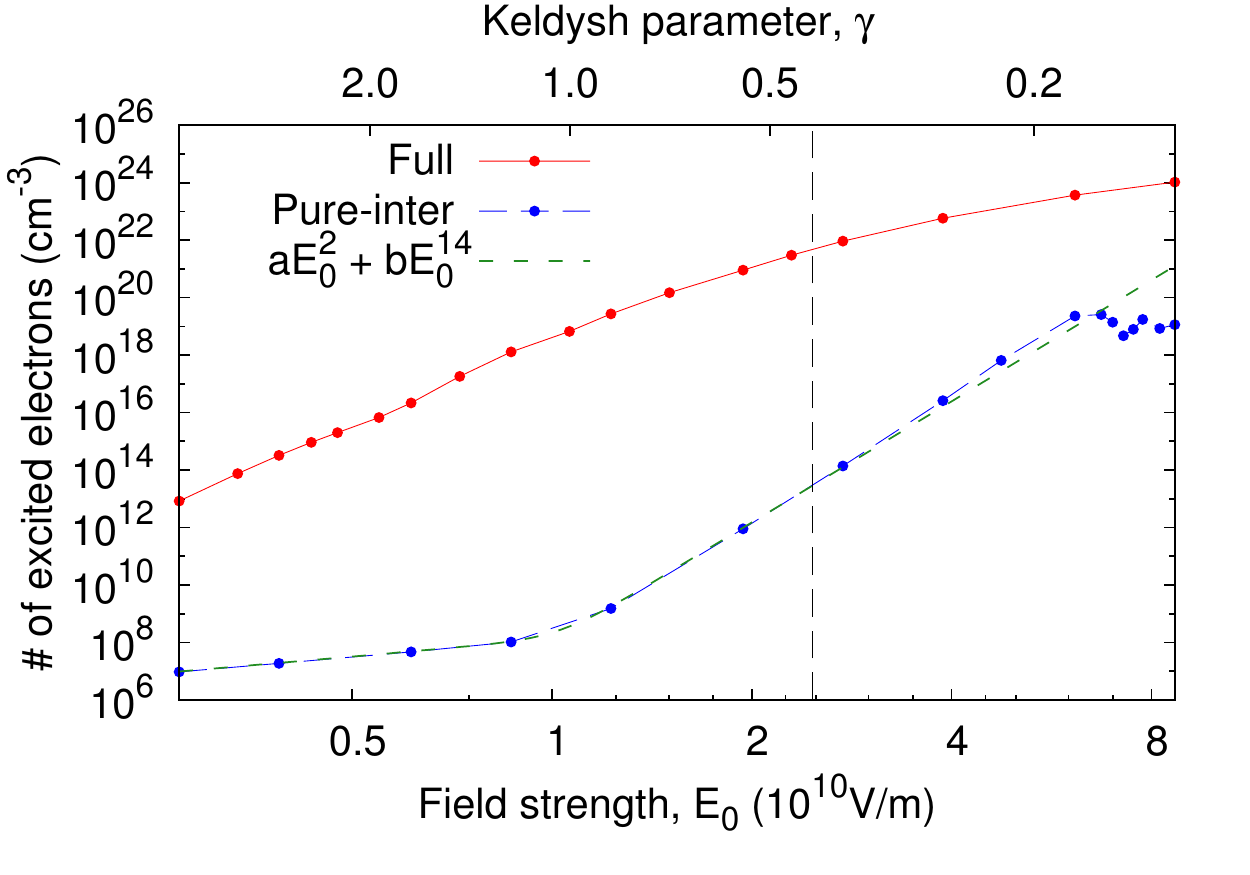}
\caption{\label{fig:nex_w155ev} Number of excited electron-hole pairs
after the laser irradiation in the strongly off-resonant regime,
computed using Eq. (\ref{eq:nex_houston}).
The dots represent the result of the full model (red solid), 
and the pure inter-band model (blue dashed).
A polynomial fitting by $aE^2_0+bE^{14}_0$ is also shown as
a green-dotted line.
}
\end{figure}

Figure \ref{fig:nex_w155ev} shows the number of excited electrons after 
laser irradiation, computed by the following formula:
\be
n_{ex} = \frac{2n_c}{(2\pi)^3}\int d\vec k  |c_{c\vec k}(t_f)|^2.
\label{eq:nex_houston}
\ee
This expression corresponds to the population in the conduction band computed
by the projection onto the conduction Houston states $u^H_{c\vec k}(\vec r,t)$.

In the figure, the red points show the result of the full model,
and the blue points show that of the pure inter-band model.
As expected, one sees that photo-carrier generation is significantly 
affected by the omission of intra-band transitions in the whole investigated
intensity region and as a consequence, photo-carrier generation is strongly suppressed in the pure inter-band case.

Here, we found that the number of photo-excited carriers of the pure inter-band model
can be well described by a simple polynomial as a function of the field strength 
$n_{ex}(E_0) = \alpha E^2_0 + \beta E^{14}_0$ even in the high intensity region.
The first term corresponds to photo-carrier generation by single-photon 
absorption due to the high-energy tail of the pulse spectrum, while 
the second term corresponds to the absorption of seven photons.
Note that, since even-number multi-photon absorption processes
are forbidden in the pure inter-band model,
the seven-photon absorption process is the lowest order nonlinear excitation process
even though the energy of six photons exceeds the gap.

This fact indicates that photo-carrier generation without intra-band transitions
is dominated by the multi-photon absorption process even at high intensities,
where the Keldysh parameter $\gamma$ is of the order of 1 or less. Note that, the Keldysh 
parameter $\gamma$ becomes one when the maximum field $E_0$ is about $10^{10}$ V/m
in the present condition (see the secondary $x$-axis of Fig.~\ref{fig:nex_w155ev}).
One can understand the predominance of the multi-photon absorption process 
in the pure inter-band model based on the ponderomotive energy. As mentioned above, 
the ponderomotive energy originates from the quiver motion due to intra-band motion.
Thus, the omission of intra-band transitions corresponds to setting 
the ponderomotive energy $U_p$ to zero, and the Keldysh parameter $\gamma$ becomes infinity. 
Therefore, one expects the excitation process to be dominated
by multi-photon absorption once intra-band transitions are neglected.

To obtain further insight into the role of intra-band transitions,
we then calculate the energy distribution of electron-hole (e-h) pairs 
at each e-h excitation energy, $n_{e-h}(E_{e-h})$.
Since we consider the two-band model in this work, 
the distribution of e-h pairs at each e-h excitation energy $E_{e-h}$ 
can be evaluated by
\be
n_{e-h}(E_{e-h}) = \frac{2n_c}{(2\pi)^3}\int d\vec k  |c_{c\vec k}(t_f)|^2
\delta (\epsilon_{c\vec k} - \epsilon_{v\vec k} - E_{e-h}). \nonumber\\
\label{eq:nex_houston_dist}
\ee

\begin{figure}[htbp]
\centering
\includegraphics[width=0.9\columnwidth]{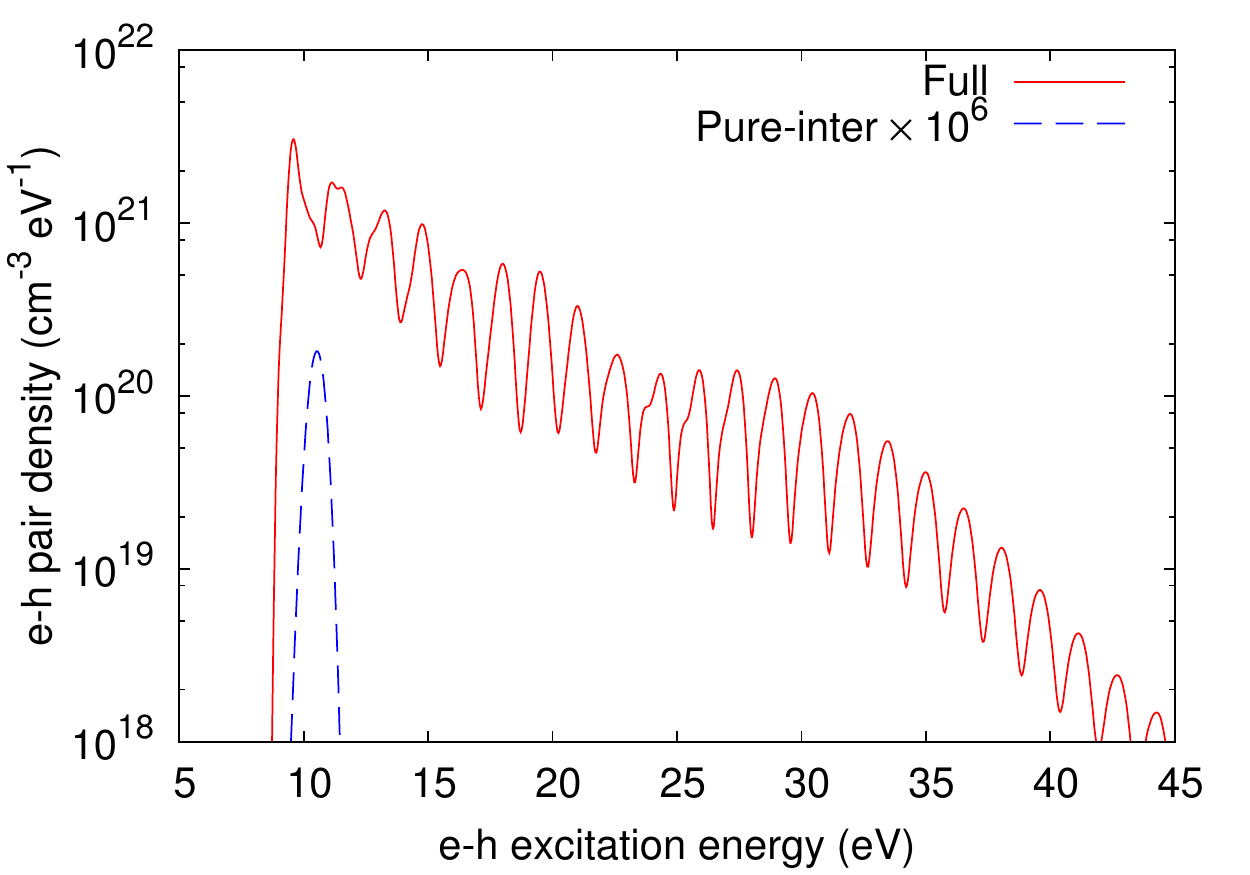}
\caption{\label{fig:nph_dist_w155ev} 
Distribution of photo-induced electron-hole pairs
in the strongly off-resonant condition ($\hbar\omega =1.55$~eV) with 
the field strength of $2.74\times10^{10}$~V/m. 
The results of the full model (red solid)
and the pure inter-band model (blue dashed) are shown.
}
\end{figure}

Figure \ref{fig:nph_dist_w155ev} shows the distribution of
e-h pairs as a function of the e-h energy after laser irradiation with a field strength
of $E_0=2.74\times 10^{10}$~V/m, which is marked with the black-dashed vertical line
in Fig.~\ref{fig:nex_w155ev}. 
The red-solid line shows
the result of the full model, while the blue-dashed line shows 
that of the pure inter-band model. 
One sees that the e-h distribution of the pure inter-band model shows 
a single peak. As discussed above,
the carrier injection in the pure inter-band model at the present
field strength is dominated by the seven-photon
absorption process. This is also reflected in the position of the single peak of the pure inter-band model, which corresponds to the energy of the seven absorbed photons
($1.55$ eV $\times 7 = 10.85$ eV).
In contrast, the e-h distribution of the full model shows multiple peaks
that are distributed over a broad energy range. 
Furthermore, one may see a plateau around $25-30$ eV.
Such a multi-peak structure with a plateau region can also be found in 
above-threshold-ionization (ATI) photoelectron spectra of atoms
\cite{PhysRevLett.42.1127,PhysRevLett.70.1599,EBERLY1991331}.
The plateau structure of the ATI spectra is formed by scattering 
of ionized electrons from parent ions under laser fields \cite{0953-4075-27-21-003}.
This indicates that the plateau structure in the e-h distribution 
of Fig. \ref{fig:nph_dist_w155ev} is also formed by some scattering process
driven by the strong electric fields.

The cut-off energy of the plateau structure of 
the ATI spectra as well as the high-order harmonic
generation from atoms has been well understood based on semi-classical models
\cite{PhysRevLett.71.1994,0953-4075-27-21-003,PhysRevA.49.2117}.
Recently, the semi-classical model for high-order harmonic generation
was extended also for solid-state systems \cite{PhysRevB.91.064302}.
Likewise, it will be important to construct a classical analog to
correctly understand the plateau structure in Fig. \ref{fig:nph_dist_w155ev}
as well as to develop an intuitive description of photo-carrier injection
under strong fields.
However, since the detailed analysis of the plateau structure based on 
a semi-classical treatment is beyond the scope of the present work, 
this aspect will be investigated in future work.

We note that the detailed structure of the e-h distribution 
must be immediately smeared out by electron-electron, 
electron-phonon, as well as electron-impurity scattering processes
in a realistic system.
Therefore, it would be hard to experimentally observe the e-h distribution
shown in Fig. \ref{fig:nph_dist_w155ev}.

\subsection{Multi-photon resonant condition; $\hbar \omega /\epsilon_g = 1/3$ 
\label{subsec:three-photon-resonance}}

Next, we investigate the effect of intra-band transitions in the multi-photon
resonant condition. For this purpose, we set the mean photon energy
of the laser pulse $\hbar \omega$ to $3$~eV, which is identical to
one third of the band gap $\epsilon_g$. As a result, 
resonant excitation is obtained by absorbing three photons.
The full pulse duration is set to $T=30$~fs.

\begin{figure}[htbp]
\centering
\includegraphics[width=0.9\columnwidth]{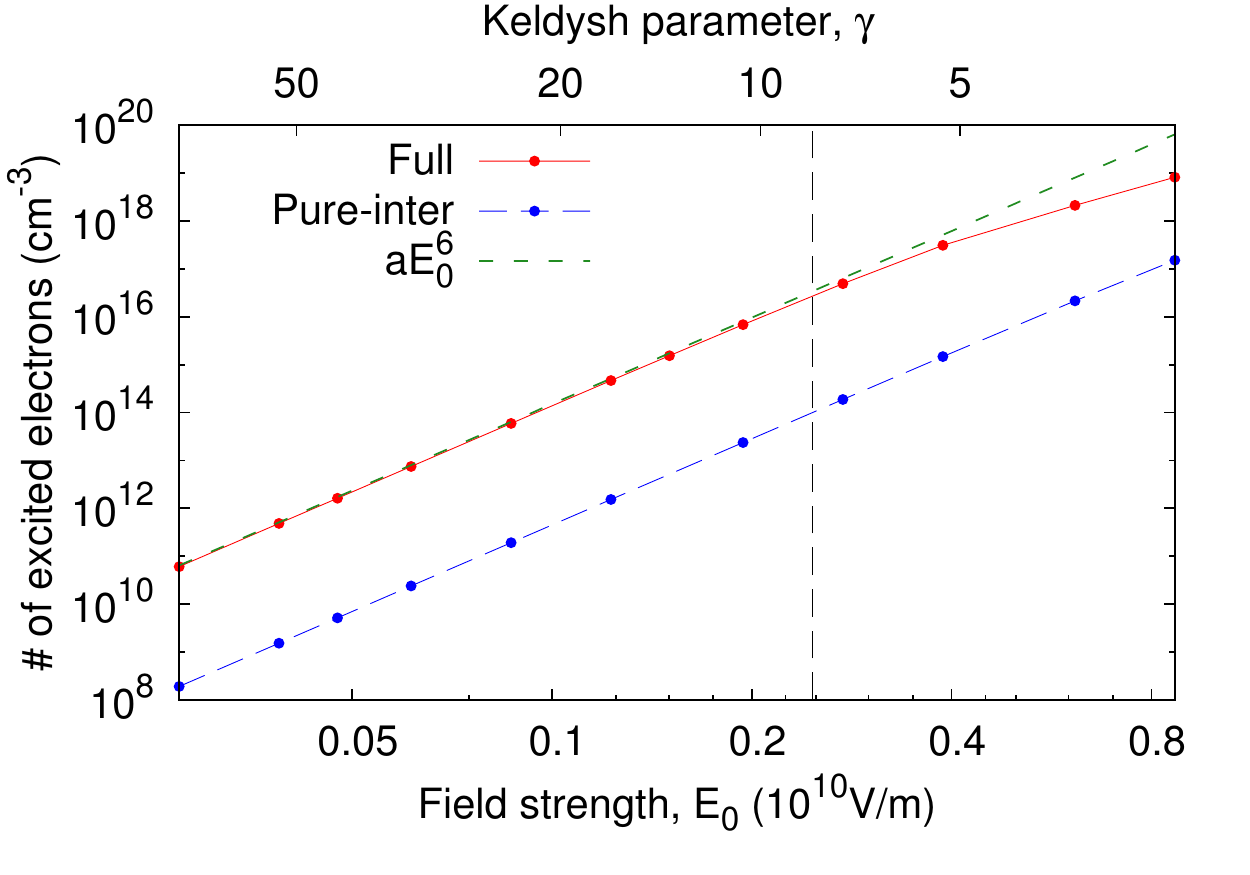}
\caption{\label{fig:nex_w300ev} 
Number of excited electron-hole pairs
after the laser irradiation in the three-photon resonant regime.
The dots represent the result of the full model (red solid), 
and the pure inter-band model (blue dashed).
A polynomial fitting by $aE^6_0$ is also shown as a green-dotted line.
}
\end{figure}

Figure \ref{fig:nex_w300ev} shows the injected carrier population
as a function of the laser field strength $E_0$ in the three-photon resonant condition.
The sixth power of the field strength, $E^6_0$, is also shown as a green
dotted line. As seen from the figure, the population of both the full model
and the pure inter-band model is proportional to the sixth power of the field strength
in the weak field region, indicating that photo-carrier injection is dominated by three-photon absorption.
Although the three-photon absorption process dominates the carrier injection 
in both models, the full model shows about $300$ times higher injected population than the pure inter-band model.
This fact means that intra-band transitions strongly enhance the multi-photon
excitation process.

Since intra-band transitions cannot directly inject any carriers
by themselves, the enhancement can be understood as photo-assisted carrier-injection
via additional excitation paths opened by intra-band transitions.
Note that, if intra-band transitions are induced by a static electric field and 
inter-band transitions are induced by an optical field,
carrier-injection by optical absorption corresponds
to photo-assisted tunneling,
which is also known as the Franz-Keldysh effect \cite{franz1958,keldysh1958}.

\begin{figure}[htbp]
\centering
\includegraphics[width=0.9\columnwidth]{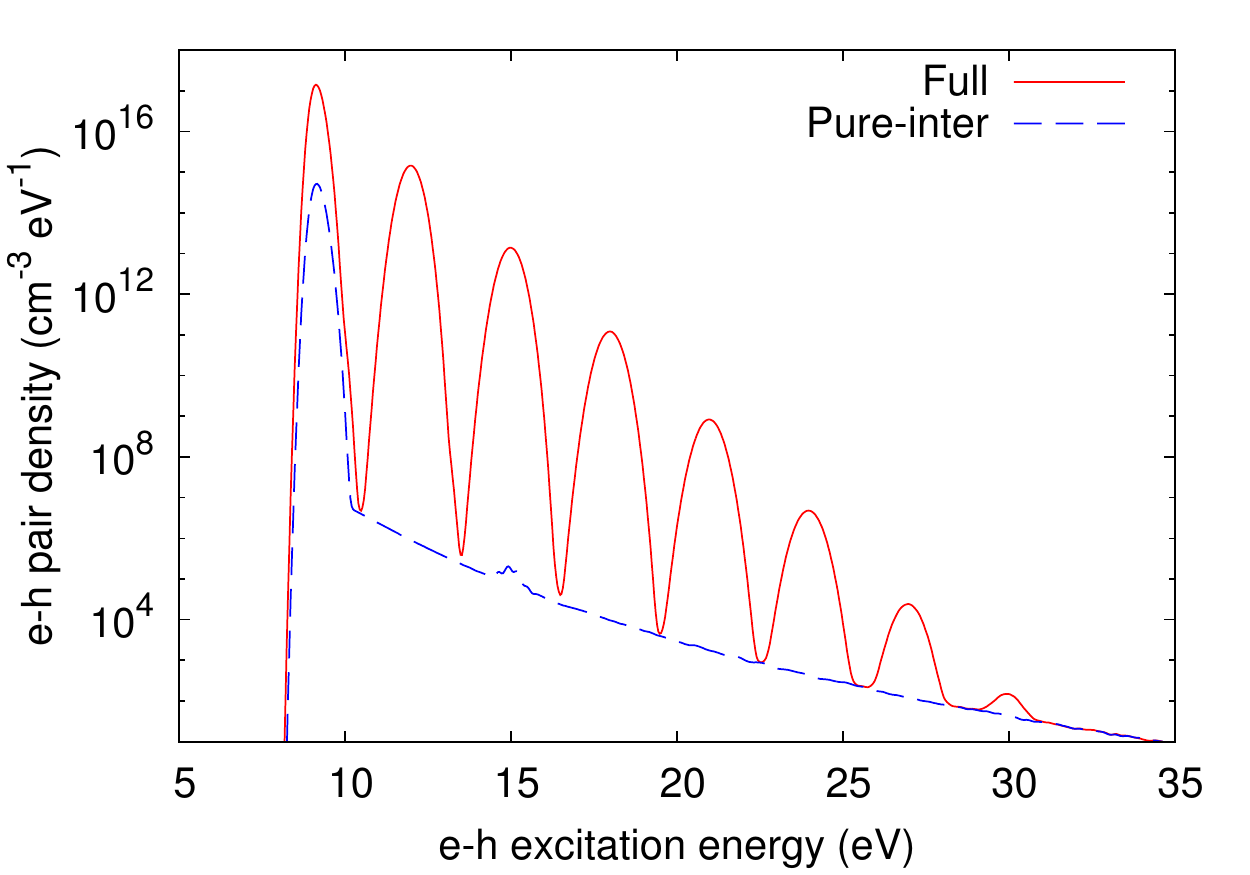}
\caption{\label{fig:nph_dist_w300ev} Distribution of photo-induced electron-hole pairs
in the three-photon resonant condition with the field strength of 
$0.274\times10^{10}$~V/m. The results of the full model (red solid)
and the pure inter-band model (blue dashed) are shown.
}
\end{figure}

Figure \ref{fig:nph_dist_w300ev} shows the distribution of e-h pairs
after laser irradiation with a field strength of $E_0=0.274\times10^{10}$~V/m,
which is marked with the black-dashed vertical line in Fig.~\ref{fig:nex_w300ev}. 
The red-solid line shows the result of the full model, while the blue-dashed line
shows that of the pure inter-band model.
As seen from the figure, the first peak at around $9$~eV
dominates the total population in both models. 
This fact is consistent with the above finding
that the photo-carrier injection is dominated by the three-photon absorption process.
Furthermore, the first peak is strongly enhanced by intra-band transitions.
Therefore, we can clearly conclude that intra-band transitions
strongly assist the lowest-order multi-photon absorption process
and largely enhance photo-carrier injection.

In Fig. \ref{fig:nph_dist_w300ev}, one sees that intra-band transitions
induce additional peaks in the higher energy region.
For the present field strength $E_0=0.274\times10^{10}$~V/m,
contributions from these additional peaks are rather small compared with 
the primary peak at around $9$~eV.
However, once the field strength becomes large enough and the corresponding
Keldysh parameter becomes small enough, 
these additional peaks are important as discussed in 
the subsection \ref{subsec:strong-off-resonant}.

To obtain further insight into the carrier-injection enhancement
by intra-band transitions, we analytically study photo-carrier generation
based on a perturbation theory. For simplicity, we consider
the carrier-injection only at the $\Gamma$-point ($\vec k=0$), assuming that the laser pulse has 
a rectangular envelope with duration $T$. 
Based on the third order perturbation theory, the injected population of the pure inter-band model
is given by
\be
n^{pure-inter}_{\Gamma} = \left | 
\frac{1}{8} \left (\frac{\vec p_{vc}\cdot \vec E_0}{\epsilon_g} \frac{e\hbar}{m} \right )
\frac{15}{16} \frac{1}{\hbar^2\omega} \Delta_S T
\right |^2,
\label{eq:3ph-inter}
\ee
where $\Delta_S$ is the cycle-averaged Stark shift induced by inter-band transitions;
\be
\Delta_S= \left (\frac{\vec p_{vc}\cdot \vec E_0}{\epsilon_g} \frac{e\hbar}{m} \right )^2\frac{1}{\epsilon_g}.
\label{eq:stark-shift}
\ee

In the full model, intra-band transitions open an additional excitation path.
The injected population via the additional path can be evaluated by
\be
n^{intra-assisted}_{\Gamma} = \left |
\frac{1}{8} \left ( \frac{\vec p_{vc}\cdot \vec E_0}{\epsilon_g} \frac{e\hbar}{m}  \right )
\frac{1}{\hbar^2 \omega} U_p T
\right |^2.
\label{eq:3ph-intra}
\ee
Detailed derivation of Eq.~(\ref{eq:3ph-inter}) and Eq.~(\ref{eq:3ph-intra})
will be described in 
Appendix~\ref{appendix:three-photon}.

With the present parameterization, the injected carrier population via the additional path, 
$n^{intra-assisted}_{\Gamma}$, is about $100$ times larger than that 
of the pure inter-band path, $n^{pure-inter}$.
This fact is consistent with the above enhancement by intra-band transitions
in the numerical simulations.

Comparing Eq. (\ref{eq:3ph-inter}) and Eq. (\ref{eq:3ph-intra}),
one sees that the ratio of the injected population via two different excitation paths
can be well approximated by the square of the ratio of
the ponderomotive energy and the Stark shift;
\be
n^{intra-assisted}_{\Gamma}/n^{pure-inter}_{\Gamma} \approx |U_p/\Delta_S|^2.
\label{eq:ratio-intra-inter-3ph}
\ee

Since only the ponderomotive energy depends on the photon energy of laser fields,
Eq.~(\ref{eq:ratio-intra-inter-3ph}) indicates that the additional 
excitation path, which is opened by intra-band transitions, becomes
significant in the strongly off-resonant regime, where the photon energy becomes
small. Furthermore, in higher order nonlinear responses,
the higher order ratio, $\left (U_p/\Delta_S \right)^n$, is expected
to be a characteristic parameter instead of Eq. (\ref{eq:ratio-intra-inter-3ph}). Therefore, once the pondermotive energy
becomes substantially large, the higher order nonlinear responses
can be strongly suppressed by the omission of intra-band transitions.
In fact, these expectations are consistent with above findings from 
the numerical simulations:
the omission of intra-band transitions strongly suppresses photo-carrier generation
in the off-resonant condition,
as seen from Fig.~\ref{fig:nex_w155ev}. Moreover, the higher order 
multi-photon peaks are strongly suppressed by the omission of intra-band transitions,
as seen from Fig.~\ref{fig:nph_dist_w155ev} and Fig.~\ref{fig:nph_dist_w300ev}.

Combining the findings in this subsection with those 
in the subsection \ref{subsec:strong-off-resonant},
we can conclude that intra-band transitions
play a crucial role in nonlinear photo-carrier injection in both multi-photon and tunneling regimes.

\subsection{Resonant condition; $\hbar \omega /\epsilon_g = 1$}

Finally, we investigate the effect of intra-band transitions in the resonant case.
For the sake of the investigation, we set the mean photon energy of the laser 
pulse $\hbar \omega$ to $9$~eV, which is identical to the band gap $\epsilon_g$.
Therefore, single-photon absorption dominates the carrier generation
in the weak-field regime.
The full pulse duration is set to $T=30$~fs.

\begin{figure}[htbp]
\centering
\includegraphics[width=0.9\columnwidth]{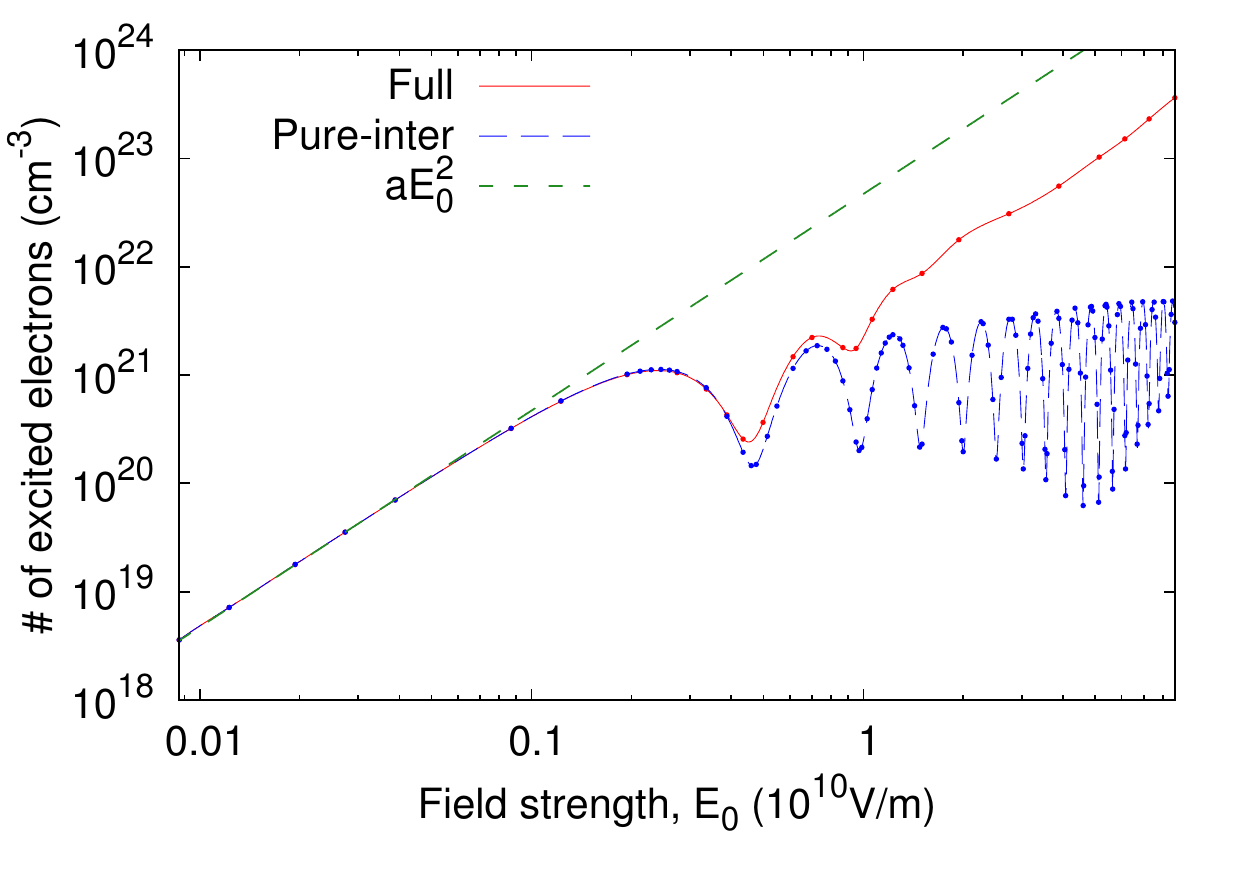}
\caption{\label{fig:nex_w900ev}
Number of excited electron-hole pairs
after the laser irradiation in the resonant regime.
The dots represent the result of the full model (red solid), 
and the pure inter-band model (blue dashed).
The square of the field strength, $E^2_0$, is also shown as
the green dotted line.
}
\end{figure}

Figure \ref{fig:nex_w900ev} shows the carrier population as a function
of the laser field strength.
One sees that the result of the full model coincides with that of
the pure inter-band model in the weak-field region.
This fact means that intra-band transitions do not play any role
in the weak-field resonant condition.
Around a field strength of $0.3\times10^{10}$~V/m, the number of excited electrons
decreases as the intensity increases. This feature indicates depopulation of excited electrons through the onset of Rabi-flopping.
In the high intensity region, the two models show different behaviors: While 
the population of the pure inter-band model shows oscillations and does not 
increase much, 
that of the full model monotonically increases. The oscillating behavior
in the pure inter-band model can be explained by the population and depopulation 
dynamics of the Rabi flopping.
The monotonic increase in the full model means that 
intra-band transitions strongly enhance photo-carrier generation and overcome the depopulation mechanism of Rabi-flopping.

\begin{figure}[htbp]
\centering
\includegraphics[width=0.9\columnwidth]{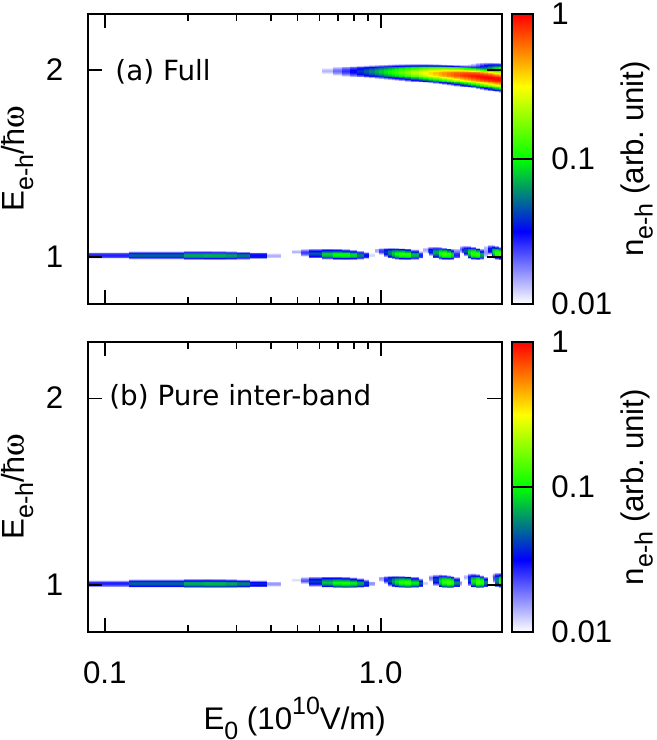}
\caption{\label{fig:nph_dist_resonant} Distribution 
of photo-induced electron-hole pairs in the resonant condition.
The result of the full model shown in (a), while that of
the pure inter-band model is shown in (b).
}
\end{figure}

In order to elucidate the microscopic origin of the enhancement of
carrier injection in the resonant condition,
we investigate the e-h pair distribution after the laser irradiation.
Figure \ref{fig:nph_dist_resonant} shows the e-h pair distribution as a function
of the e-h excitation energy $E_{e-h}$ and the field strength $E_0$.
Figure \ref{fig:nph_dist_resonant} (a) shows the result of the full model,
while Fig. \ref{fig:nph_dist_resonant} (b) shows that of the pure inter-band model.
As seen from Fig. \ref{fig:nph_dist_resonant} (b), the e-h pairs are created
at around the photon energy of the laser field ($E_{e-h}\sim \hbar \omega$) 
in the entire investigated
field strength range. Therefore, photo-carrier injection
is always dominated by single-photon absorption in the pure inter-band model.
Furthermore, one may see that the population of e-h pairs at around
the photon energy, $E_{e-h}\sim \hbar \omega$, in the pure inter-band model
shows an oscillating behavior with increasing field strength $E_0$.
This oscillating behavior corresponds to the oscillations observed in Fig. \ref{fig:nex_w900ev},
and it originates from Rabi flopping.
In Fig. \ref{fig:nph_dist_resonant} (a), a similar structure can be found 
at the photon energy of the laser field, $E_{e-h} = \hbar \omega$.
This fact indicates that intra-band transitions do not directly affect
the Rabi flopping dynamics and the single-photon absorption. 
One sees that intra-band transitions induce additional large numbers of e-h pairs at around $E_{e-h}=2 \hbar \omega$.
This fact indicates that intra-band transitions open a multi-photon excitation 
channel and strongly enhance the carrier-injection once the field strength becomes 
strong enough.

\section{Summary \label{sec:summary}}

In this work, we investigated the role of intra-band transitions
in photo-carrier generation in semiconductors and insulators 
based on the parabolic two-band model.
The accuracy of the model was demonstrated by comparing it with the non-parabolic 
Kane's band model and \textit{ab-initio} TDDFT simulations.

We first studied the photo-carrier generation under the off-resonant condition,
where the mean photon energy $\hbar \omega$ is much smaller than the band gap
$\epsilon_g$. We found that, in this off-resonant regime, 
intra-band transitions are indispensable for the description
of photo-carrier generation.
Furthermore, if intra-band transitions are not taken into account, 
the injection mechanism is dominated by multi-photon absorption
even for high laser intensities. This fact can be understood
based on the Keldysh parameter $\gamma=\sqrt{\epsilon/4U_p}$:
The ponderomotive energy $U_p$ is induced by the quiver motion in the momentum space, 
which is an intra-band mechanism. Once intra-band transitions are ignored, the effective 
ponderomotive energy vanishes, and the Keldysh parameter becomes infinity.
Thus, the excitation mechanism is dominated by the multi-photon process if
intra-band transitions are ignored.
Moreover, we investigated the electron-hole (e-h) pair distribution
as a function of the e-h excitation energy.
As a result, we found that the e-h pair distribution under 
a strong field shows a multi-peak structure with a plateau region
(see Fig. \ref{fig:nph_dist_w155ev}).
A similar feature has been found in 
above-threshold-ionization (ATI) 
photoelectron spectra of atoms. The origin of the plateau of the ATI spectra
has been understood as the scattering of ionized electrons from the parent ion
based on the semi-classical model \cite{0953-4075-27-21-003}.
Therefore, the formation of the plateau in the e-h pair distribution might 
also be explained by a semi-classical description.

We then investigated photo-carrier generation in the three-photon
resonant condition, where the mean photon-energy is identical
to one third of the band gap.
We found that intra-band transitions largely enhance
the three-photon absorption process.
To clarify the origin of the enhancement, we analytically studied
the three-photon absorption based on the perturbation theory.
As a result, we found that intra-band transitions
open additional excitation paths that generate photo-carriers much more efficiently than the pure inter-band excitation path.
We note that, if intra-band transitions are induced by a static electric field,
the mechanism corresponds to photo-assisted tunneling, or the
so-called Franz-Keldysh effect \cite{franz1958,keldysh1958}.

We then investigated the carrier-injection in the resonant condition, where
the mean photon-energy of the laser field is identical to the band gap.
When the field is weak enough, the carrier-injection is dominated by
single-photon absorption, and intra-band transitions do not play
any role. In contrast, once the field becomes strong enough, 
intra-band transitions significantly enhance the photo-carrier generation.
Based on the energy-resolved e-h distribution analysis, 
we clarified that the enhancement of the photo-carrier generation originates from 
additional multi-photon excitation paths opened by intra-band transitions.

Starting from the above analysis, 
we can conclude that intra-band transitions largely enhance
the carrier-injection once nonlinear effects become substantial.
This finding indicates a potential to control photo-carrier injection
by employing multi-color laser pulses: some of the pulses mainly induce 
the carrier-injection via inter-band transitions, while the others assist
it by opening efficient excitation paths via intra-band transitions.
Here, in addition to the photon-energy of each pulse, 
the pulse width, the relative time delay, and the relative carrier-envelope phase can be optimizable
parameters.
An efficient enhancement or suppression of the carrier-injection
by optical laser pulses will be important for technological applications such as light-driven control of material properties as well as for fundamental investigations of electron dynamics in solids.

\section*{Acknowledgements}

This work was supported by the European Research Council (ERC-2015-AdG-694097), 
Grupos Consolidados (IT578-13),
European Union's H2020  programme under GA no.676580 (NOMAD) and
National Center of Competence in Research Molecular Ultrafast Science and Technology (NCCR MUST) funded by the Swiss National Science Foundation.
S.A.S gratefully acknowledges the fellowship by the Alexander von Humboldt Foundation.

\appendix
\section{Perturbation analysis for three-photon absorption process
\label{appendix:three-photon}}

Here, we describe the detailed derivation of Eq.~(\ref{eq:3ph-inter}) and
Eq.~(\ref{eq:3ph-intra}), analyzing the tree-photon absorption process
with the perturbation theory.

First, we analyze the pure inter-band model, ignoring
intra-band transitions.
At the $\Gamma$-point ($\vec k=0$), the Schr\"odinger equation (\ref{eq:schrodinger-2band})
of the pure inter-band model can be described as
\be
i\hbar \dot {\bf c}(t) = H_{inter}(t) {\bf c}(t),
\label{eq:inter-schrodinger}
\ee
\be
H_{inter}(t) =
\left(
    \begin{array}{cc}
      0 & h_{inter}(t)  \\
      h^*_{inter}(t) & 0
    \end{array}
  \right),
\ee
and 
\be
h_{inter}(t) = i\frac{ \vec p_{vc} \cdot \vec E(t)} 
{\epsilon_g}\frac{e \hbar}{m}
e^{\frac{1}{i\hbar} \epsilon_g t},
\ee
where ${\bf c}(t)$ is a two-dimensional row vector.

Since $H_{inter}(t)$ contains only the first order term of $\vec E_0$,
one can simply consider the following perturbation expansion,
\be
{\bf c}(t) = {\bf c}^{(0)}(t) + {\bf c}^{(1)}(t)
+ {\bf c}^{(2)}(t) + {\bf c}^{(3)}(t) + \cdots.
\label{eq:perturbation-expansion}
\ee
Inserting Eq.~(\ref{eq:perturbation-expansion}) into Eq.~(\ref{eq:inter-schrodinger}),
each perturbation order can be evaluated recursively;
\be
{\bf c}^{(1)}(t) = \frac{1}{i\hbar} \int^t_0 dt' H_{inter}(t'){\bf c}^{(0)}(t'),
\label{eq:1st-pert}
\ee
\be
{\bf c}^{(2)}(t) = \frac{1}{i\hbar} \int^t_0 dt' H_{inter}(t'){\bf c}^{(1)}(t'),
\ee
\be
{\bf c}^{(3)}(t) = \frac{1}{i\hbar} \int^t_0 dt' H_{inter}(t'){\bf c}^{(2)}(t'),
\label{eq:3rd-pert}
\ee
and so on.

To simply evaluate the above perturbation expansion,
we assume the three-photon resonant condition ($3\hbar \omega = \epsilon_g$)
and a rectangular envelope with duration $T$
for the applied laser field.
Thus, the electric field can be described by
\be
\vec E(t) = \vec E_0 \cos(\omega t),
\label{eq:elec-field}
\ee
in the domain $0<t<T$ and zero outside.
Further assuming that the wavefunction is initially set to the ground state,
the third order coefficient after the laser irradiation
can be evaluated as
\be
{\bf c}^{(3)}(t>T) &=& \frac{1}{(i\hbar)^3} \int^T_0 dt' 
\int^{t'}_0 dt'' 
\int^{t''}_0 dt''' 
H_{inter}(t') \nonumber \\
&&
\times H_{inter}(t'')H_{inter}(t''') 
\left(
    \begin{array}{c}
      1   \\
      0
    \end{array}
  \right).
\ee
Ignoring oscillatory integrands for $t'$, the conduction-band component of
${\bf c}^{(3)}(t>T)$ can be simply expressed as
\be
c^{(3)}_c(t>T) &=& - \frac{1}{8} \left 
(\frac{\vec p_{vc}\cdot \vec E_0}{\epsilon_g} \frac{e\hbar}{m} 
\right )^{3}
\frac{5}{16\hbar^3\omega^2}T \nonumber \\
&=& - \frac{1}{8} \left 
(\frac{\vec p_{vc}\cdot \vec E_0}{\epsilon_g} \frac{e\hbar}{m} 
\right )
\frac{15}{16\hbar^2\omega} \Delta_S T,
\ee
where $\Delta_S$ is the cycle-averaged Stark shift defined in Eq.~(\ref{eq:stark-shift}).
In the last line, we employed the three-photon resonant condition;
$3\hbar \omega = \epsilon_g$.
Finally, the injected population is given by
\be
n^{pure-inter}_{\Gamma} &=& |c_c(t>T)|^2 \nonumber  \\
&=& \left | 
\frac{1}{8} \left (\frac{\vec p_{vc}\cdot \vec E_0}{\epsilon_g} \frac{e\hbar}{m} \right )
\frac{15}{16} \frac{1}{\hbar^2\omega} \Delta_S T
\right |^2. \nonumber \\
\ee

Note that the first and the second order contributions, ${\bf c}^{(1)}(t)$ and ${\bf c}^{(2)}(t)$,
become zero after the laser irradiation in the pure inter-band model.

Then, we investigate the effect of intra-band transitions.
Although the Hamiltonian of the pure inter-band model, $H_{inter}$, contains only 
the first order perturbation,
that of the full model may contain higher order perturbations
due to intra-band transitions. Therefore, intra-band transitions open 
additional excitation paths.

To take intra-band transitions into account, 
we replace the Hamiltonian $H_{inter}$ by
\be
H(t) =
\left(
    \begin{array}{cc}
      0 & h(t)  \\
      h^*(t) & 0
    \end{array}
  \right),
\ee
and 
\be
h(t) = i\frac{ \vec p_{vc} \cdot \vec E(t)} 
{\epsilon_g}\frac{e \hbar}{m}
e^{\frac{1}{i\hbar} \epsilon_g t
+\frac{1}{i\hbar}\int^t_0 dt'\frac{e^2 A^2(t')}{2m_r c^2}}.
\label{eq:appendix-off-diagonal}
\ee
Here, the effect of intra-band transitions is included 
via the vector potential $A(t)$. 
We note that we simply ignored the time-dependence of the denominator in $h(t)$.
According to Eq.~(\ref{eq:elec-field}), the vector potential has the following form
\be
\vec A(t) = -c \frac{\vec E_0}{\omega} \sin(\omega t),
\ee
in the domain $0<t<T$ and zero outside. 

Assuming the small amplitude for the electric field $\vec E_0$,
the off-diagonal element of Eq.~(\ref{eq:appendix-off-diagonal}) can be approximated by
\be
h(t) = i\frac{ \vec p_{vc} \cdot \vec E(t)} 
{\epsilon_g}\frac{e \hbar}{m}
e^{\frac{1}{i\hbar} \epsilon_g t}
\left ( 1 +\frac{1}{i\hbar}\int^t_0 dt'\frac{e^2 A^2(t')}{2m_r c^2}
\right ). \nonumber \\
\label{eq:appendix-off-diagonal-exp}
\ee

Then, we evaluate Eq.~(\ref{eq:1st-pert}) by replacing $H_{inter}(t)$ with $H(t)$.
Although ${\bf c}^{(1)}(t)$ became zero after the laser irradiation in the pure
inter-band model, here it can be non-zero due
to the higher order contribution in Eq.~(\ref{eq:appendix-off-diagonal-exp}).
Ignoring oscillatory integrands in Eq.~(\ref{eq:1st-pert}), 
one can easily evaluate the conduction component of ${\bf c}^{(1)}(t)$
with $H(t)$ as
\be
c^{(1)}_c(t>T)= -\frac{1}{8}
\left 
(\frac{\vec p_{vc}\cdot \vec E_0}{\epsilon_g} \frac{e\hbar}{m} 
\right ) \frac{1}{\hbar^2\omega}U_p T,
\ee
and hence
\be
n^{intra-assisted}_{\Gamma} &=& |c^{(1)}_c(t>T)|^2 \nonumber  \\
&=&\left |
\frac{1}{8} \left ( \frac{\vec p_{vc}\cdot \vec E_0}{\epsilon_g} \frac{e\hbar}{m}  \right )
\frac{1}{\hbar^2 \omega} U_p T
\right |^2. \nonumber \\
\ee

\bibliographystyle{apsrev4-1}
\bibliography{ref}

\end{document}